# Heat localization through reduced dimensionality

Mike Chang,[1,4] Harrison D. E. Fan,[1,4] , Mokter M. Chowdhury, [1,4] George A. Sawatzky,[2,3,4] Alireza Nojeh[1,4*]

[1]Department of Electrical and Computer Engineering, University of British Columbia, Vancouver, British Columbia
V6T 1Z4, Canada
[2]Department of Physics and Astronomy, University of British Columbia, Vancouver, British Columbia
V6T 1Z1, Canada
[3]Department of Chemistry, University of British Columbia, Vancouver, British Columbia V6T 1Z1, Canada
[4]Quantum Matter Institute, University of British Columbia, Vancouver, British Columbia V6T 1Z4, Canada
* Corresponding author: alireza.nojeh@ubc.ca

### Abstract

We present a model to show that heat propagation away from a local source depends strongly on dimensionality, leading to dramatic localization in low-dimensional systems. An example of such a system is a carbon nanotube array. We further show that this localization is amplified due to a runaway mechanism if thermal conductivity declines rapidly with temperature. Extremely high temperatures of thousands of kelvins and gradients of hundreds of K/µm may thus be obtained in a conductor using a modest local power source such as a laser pointer. This is of fundamental importance for high-efficiency energy conversion through thermoelectric and thermionic mechanisms, as well as various other applications.

### Introduction

At finite temperatures, electron transport leads to the transfer of a certain amount of heat as expressed by the Wiedemann-Franz law, although deviations have been observed and attributed to the lack of quasiparticles [1, 2]. The other contribution to thermal conductivity is by phonons, making it difficult to maintain a high temperature difference across a device, but this contribution can in principle be made to be small, such as in the phonon-glass structures clathrates [3] and skutterudites [4]. Indeed, limiting heat flow across a conductor or semiconductor is the key challenge in the conversion of thermal to electrical energy. Low-dimensional systems can restrict heat flow due to spatial confinement of phonons and may provide a path to address this long-standing problem [5, 6, 7]. For example, strongly localized heating and incandescence have been reported in graphene ribbons subject to an electric current [8].

Arrays of vertically-aligned multi-walled carbon nanotubes (CNT forests) are quasi one-dimensional (1D) materials with up to macroscopic dimensions, in which nanotubes have long-range alignment (Fig. 1(a)), leading to strongly anisotropic bulk properties. An extreme manifestation of heat confinement has previously been demonstrated experimentally in CNT forests [9]. In this so-called "Heat Trap" effect, a beam of light illuminates a spot on the sidewall of the CNT forest; although CNTs are good thermal conductors, surprisingly, the generated heat remains strongly localized not only in the direction perpendicular to the nanotubes, but also along the nanotubes. Figure 1(b) shows a photo of a ~1.5 mm tall CNT forest whose sidewall is illuminated with an infrared light beam focused to a point with a diameter of approximately 250 µm. The bright region corresponds to the incandescent glow of the resulting hot spot (the incident beam is not seen in the photo). Using thermographic imaging and hyperspectroscopy, temperature gradients greater than 10 K/µm have been measured along the nanotubes [10, 11]—values that are unprecedented in ordinary, isotropic bulk conductors.

This strong confinement allows the structure to reach peak temperatures of > 2,000 K (leading to thermionic electron emission following the Richardson-Dushman law [12]) with an input optical intensity on the order of only tens of W/cm$^2$, which is 3-4 orders of magnitude less than the intensities required to heat ordinary conductors to such temperatures. Such facile heating using readily-accessible optical powers, together with the fact that the effect is relatively insensitive to the illumination wavelength [13], has been exploited to create simple and compact solar thermionic electron emitters and energy converters [14]. The effect has also enabled the combination of multi-photon photo-electron emission and thermionic electron emission [15]. The physical origins of this strong heat confinement remain unknown. Adding to the puzzle is the fact that, in laser-induced heating of individual CNTs or small bundles, optical intensities thousands of times greater than those used in the Heat Trap effect have yielded peak temperatures of only a few hundred degrees [16, 17]. This suggests that the collective behavior of the large ensemble of the CNTs may play a key role in heat localization as an emergent phenomenon. In the present theoretical/numerical study, we develop a model based on anisotropic heat flow that not only provides an explanation for this effect, but may also form the foundation of an approach to confining heat while allowing electricity to flow. The key merit of this approach is that it does not rely on having an inherently low thermal conductivity, thus opening up the possibility of using materials with a variety of electrical properties for energy conversion by shaping them into 1D form.

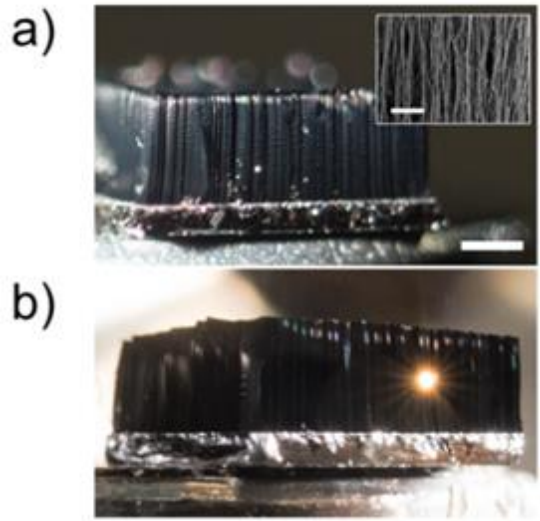

FIG. 1. "Heat Trap" on a carbon nanotube forest. (a) Close-up photograph of a ~1.5 mm-tall forest of multi-walled CNTs, showing their vertical alignment. Inset is a scanning electron micrograph of the side of a typical CNT forest. The forest was grown using chemical vapour deposition on a silicon substrate with a thin-film iron catalyst. Scale bars: 1 mm (photo) and 1 μm (inset). (b) Photo of a "Heat Trap" spot on the side of the forest under vacuum, upon illumination with an infrared laser beam (wavelength: 1064 nm; approximate spot diameter: 250 μm), which itself is not seen in the photo. Despite high conductivity along the CNTs, the temperature profile exhibits strong confinement, as seen by the localized incandescence from the illuminated, hot spot.

**The Model**

Figure 2 shows a schematic of an illuminated CNT array. Each nanotube dissipates the absorbed optical power through thermal conduction along its axis (dominated by phonons, with possible contributions from other quasiparticles and electrons), as well as re-radiation in the form of incandescence (photons) and thermionic emission (electrons). The emitted photons and electrons can in turn be partially re-absorbed by adjacent nanotubes in the forest. The average inter-nanotube spacing is a few tens of nanometers, which is much smaller than the characteristic wavelength of thermal radiation (infrared). The CNTs are thus strongly electromagnetically coupled and the far-field Stefan-Boltzmann law may not be applicable for estimating their radiative energy exchange. The full treatment of this rich and complex electrodynamic problem requires simultaneously solving the optical absorption/re-radiation, heat conduction and electron emission for a large network of coupled CNTs, while taking into account their individual and diverse electronic structures and optical and thermal characteristics. Such an all-encompassing theory is currently out of reach. Fortunately, however, a key property of the CNT forest allows for a much simpler approach. CNT forests have been reported to be extremely dark and behave like a blackbody [18, 19]; experiments have revealed that they absorb more than 99.9% of light over a broad spectral range [20]. This may be understood as being the overall result of all the re-radiation and re-absorption events by individual nanotubes, effectively leading to the capture of the entire incident optical energy by the forest. The other important consideration is related to thermal conductivity. As expected, thermal conductivity has been predicted and measured to be highly anisotropic in the CNT forest, with the value along the axis of the CNTs being one to two orders of magnitude greater than in the transverse directions [21-23]; similarly, an anisotropy of more than two orders of magnitude has also been reported in silicon nanowire arrays [24]. The question arises as to whether anisotropy persists at very high temperatures, where inter-nanotube radiative coupling, which does not depend on physical contact and phonon flow, may transfer significant amounts of heat in the transverse directions. However, measurements of the temperature profile at high temperatures have shown significantly tighter confinement in the transverse direction than along the nanotubes [10, 11], demonstrating that anisotropy persists.

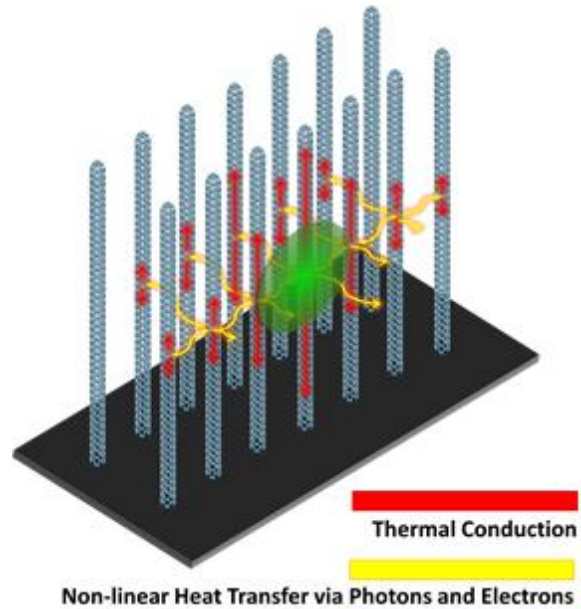

FIG. 2. Dimensionality manifesting as anisotropy. Conceptual representation of optical absorption by an array of aligned carbon nanotubes, showing thermal conduction along each, as well as radiation and electron emission to adjacent nanotubes. The green circle indicates the area to which incident power is delivered.

In the context of our model, the above observations allow us to treat the forest as a continuous bulk medium with nearly perfect optical absorption and anisotropic thermal conductivity. This is appropriate if the illuminated spot is large enough to cover many CNTs, which is the case in typical Heat Trap experiments where hundreds to thousands of CNTs are covered by the illuminating beam spot. In this context, the many individual photon and electron emission and reabsorption events within the bulk of the CNT forest lead to an effective photon emission (incandescence) and electron emission (thermionic emission) from the surface of the forest—now represented by a continuous bulk—as seen in the "Methods" section in Supplemental Material [25]. Finally, we note that thermal conductivity depends on temperature [26]. Importantly, the temperature dependence of thermal conductivity may itself be affected by dimensionality: whereas in 3D crystals thermal conductivity may peak at a temperature of a few tens of kelvins and then decline with a characteristic $\sim T^{-1}$ behavior due to Umklapp phonon scattering, in 2D and 1D materials such as graphene ribbons and carbon nanotubes, the peak may happen at hundreds of kelvins [27-31]. Therefore, for applications at room temperature and above, the role of the temperature dependence of thermal conductivity is expected to be highly pronounced in these materials.

Our model is founded on two critical observations: 1) heat flow along a given direction in a material is affected by the thermal conductivity not only in that direction, but also that in the transverse directions; 2) a decrease in thermal conductivity with temperature leads to a snowballing rise in local temperature. We use the model (see the "Methods" section in Supplemental Material [25] for details and past literature on solving the heat equation [32-40]) to calculate the temperature distribution in a material illuminated by a Gaussian optical beam with a spot diameter of 100 µm, and compare three cases: (1) isotropic-3D; (2) anisotropic-2D, where thermal conductivity along the depth is 100 times lower than along the surface directions; (3) anisotropic-1D, where thermal conductivities along the depth and one of the surface directions are 100 times lower than along the other surface direction. These are meant to conceptually represent various physical situations (a bulk, a layered structure such as multi-layer graphene, and an array of nanotubes/nanowires such as a CNT forest). To allow for a fair comparison, in each case, we consider three behaviours for thermal conductivity, all starting with a value of 100 W/mK at room temperature: (1) constant; (2) decreasing with temperature as $\sim T^{-1}$ (due to Umklapp phonon scattering [41]); and (3) decreasing with temperature primarily as $\sim T^{-2}$ at high temperatures (due to higher-order scattering) as, for example, suggested for CNTs [30]. The complete functional forms of the temperature dependences we have used are given in Supplemental Material [25]. For brevity, throughout the manuscript we refer to those as constant, $T^{-1}$ and $T^{-2}$, respectively. Note that the numerical values of the results presented below depend directly on our specific choice for the functional form of thermal conductivity; therefore, we put no emphasis on the exact values of temperature gradient or other quantities we report, but our goal is to show a qualitative comparison among the different cases of anisotropy and temperature dependence of thermal conductivity.

**Results and Discussion**

Figure 3 shows our central result: the temperature distribution on the illuminated surface. In the first row, a significantly higher peak temperature of 3,000 K is achieved for 1D (versus 1,411 K for 2D and 421 K for 3D), with a maximum temperature gradient along the longitudinal, high-conductivity direction (Y axis) of 20.4 K/µm (versus 14.8 K/µm for 2D and 1.7 K/µm for 3D). In the 1D case, not only is the temperature confined along the transverse, low-conductivity direction (X axis), but it is also highly confined along the high-conductivity axis, in sharp contrast to 3D and 2D. Limiting thermal conductivity in transverse directions thus has a decisive impact on temperature distribution along the high-conductivity axis as well. This may be viewed as a consequence of the fact that heat takes infinitely many paths in all directions in going from one point to another, and thus the different directions cannot be viewed as independent. This can be directly appreciated based on Eq. (6) in Supplemental Material [25], where it is seen that the entire spatial distribution of the so-called linear temperature (an intermediate variable in the calculation of true temperature) is inversely proportional to $(\alpha\beta\gamma^2)^{1/4}$, where $\alpha$, $\beta$, and $\gamma$ are the anisotropy factors of thermal conductivity associated with the three directions. In other words, the temperature profile in any direction is affected by conductivity in all three directions.

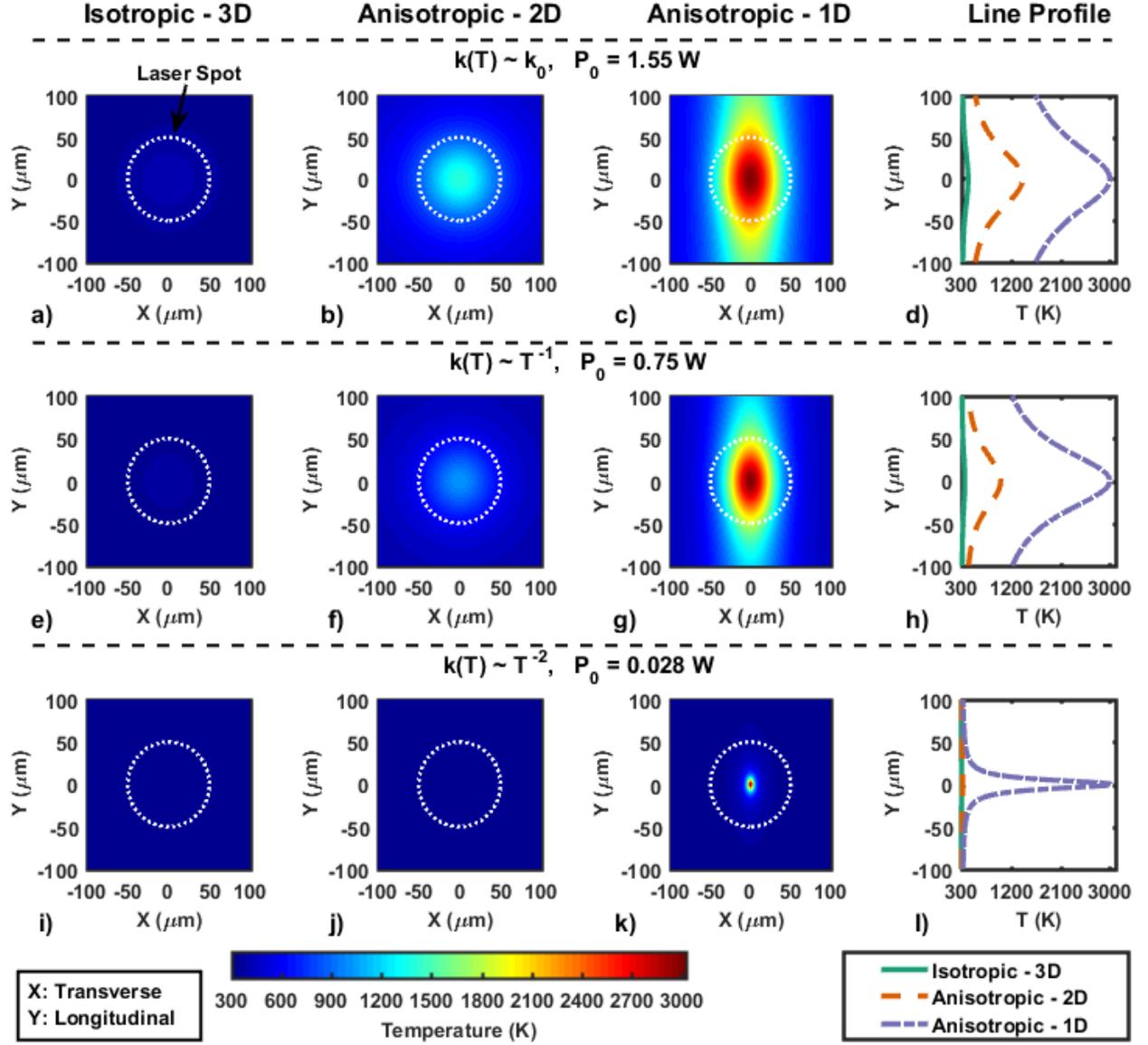

FIG. 3. Calculated temperature distribution on the illuminated surface. The first three columns from the left are for the 3D, 2D and 1D cases, respectively, and the fourth column shows the cross section of the profile along the Y axis (direction of high thermal conductivity) for all three cases. The different rows correspond to the different temperature dependences of thermal conductivity (constant, $T^{-1}$ and $T^{-2}$, respectively). For each such temperature dependence, the value of the input optical power has been chosen so as to result in a peak temperature of 3,000 K for the 1D case. The dashed circle represents the illuminating beam spot, which has an $e^{-2}$ diameter of 100 µm.

Comparing the three panels in the third column of Fig. 3, we note that the temperature localization described above is strongly amplified if thermal conductivity is a decreasing function of temperature, the effect being more severe the steeper the decrease. We interpret this as a runaway effect due to a positive feedback: a rise in temperature leads to a corresponding drop in local thermal conductivity, which in turn leads to further increase in local temperature and so on. This is as if, at high temperatures, the increase in the population of high-momentum phonons leads to a phonon traffic jam that restricts heat conduction (through Umklapp scattering). The value of total input optical power for each row of Fig. 3 has been chosen so as to obtain the same peak temperature of 3,000 K for 1D. Note that the required power has decreased by over 50 times, from 1.55 W to 0.75 W and then 0.028 W, by going from a constant thermal conductivity to one behaving as $T^{-1}$ and then $T^{-2}$. The effect of anisotropy is also most pronounced in the latter case: whereas a modest input optical power of 28 mW enables the 1D-$T^{-2}$ material to reach ~3,000 K and a maximum temperature gradient of 282.9 K/µm, it has essentially negligible effect on 2D and 3D. Similarly, this amount of input power would not lead to significant temperature increase even in 1D if the thermal conductivity did not decrease rapidly with temperature. We point out that, while the dimensionality and the temperature dependence of thermal conductivity each play a decisive role in heat flow and the resulting temperature profile, their effects are ultimately mixed and the former can exacerbate the latter, as illustrated by the extremely strong and confined heating seen in Fig. 3(k).

Figures 3(k) and 3(l) also reveal the striking feature of a hot spot that is significantly smaller than the illuminating beam spot: whereas the optical beam has an $e^{-2}$ diameter of 100 µm, the full width at half maximum (FWHM) of the resulting temperature profile is only 12.52 µm along the Y axis. Also, note that even stronger confinement and higher temperature gradients result for smaller illuminations spots. For example, for a 1 µm-diameter beam, the 1D-$T^{-2}$-conductivity material would yield a staggering maximum temperature gradient of 13,470 K/µm with 777 µW of input power, with a peak temperature of 2,987 K and an extremely confined temperature profile with a FWHM of 258 nm. (It should be noted that, at extremely high temperature gradients, a microscopic treatment of the temperature distribution on the individual CNTs involved may be necessary and Fourier's law may not apply [42], and that even the correct definition of temperature at that scale in nonequilibrium requires care [43], so the results in that regime are of qualitative value, rather than being exact predictions.)

It would also be instructive to examine the spatial distribution of thermal conductivity under localized heating. Since thermal conductivity is a tensor, instead of plotting its various components separately, and inspired by the mixing of the three components as discussed before, we have defined an effective thermal conductivity as $(\alpha\beta\gamma^2)^{1/4} k(T)$ in order to simultaneously capture the effects of anisotropy and temperature dependence. Figure 4 shows the spatial distribution of this effective thermal conductivity for all the cases of temperature distribution depicted in Fig. 3. In the first row, the effective conductivity is constant throughout space, but is reduced progressively when going from 3D to 2D to 1D. In the second row, the $T^{-1}$ behavior leads to the creation of an effectively thermally isolated area in the middle of the conductor, but the effect is negligible in the 3D case and moderate in the 2D case; it is much stronger in the 1D case. This local thermal isolation becomes much more pronounced with the $T^{-2}$ behavior as shown on the third row.

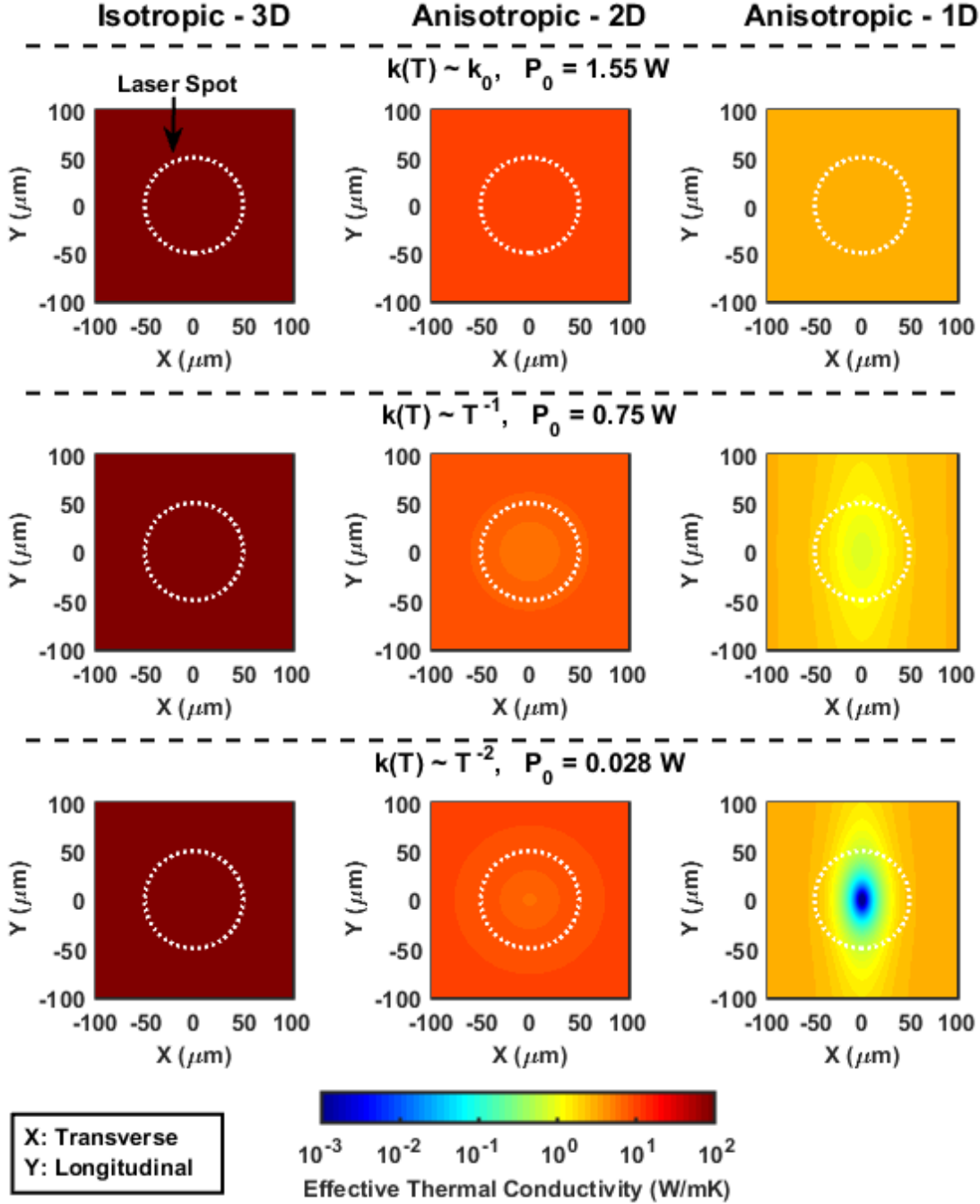

FIG. 4. Calculated effective thermal conductivity, $(\alpha\beta\gamma^2)^{1/4} \, k(T)$, on the illuminated surface corresponding to the various cases of temperature distribution shown in Fig. 3.

Figure 5 illustrates the peak temperature and the FWHM (along the high-conductivity axis) of the temperature profile versus input optical power. We see that 1D reaches high temperatures at much lower input powers than 2D and 3D. For constant thermal conductivity, the temperature rises linearly (until incandescence and thermionic emission losses become significant—not shown). For a decreasing thermal conductivity, the temperature grows nonlinearly, with the effect being significantly more pronounced for the $T^{-2}$ dependence. This effect takes over at much lower input powers for 1D, where heat transfer is severely constricted as discussed before.

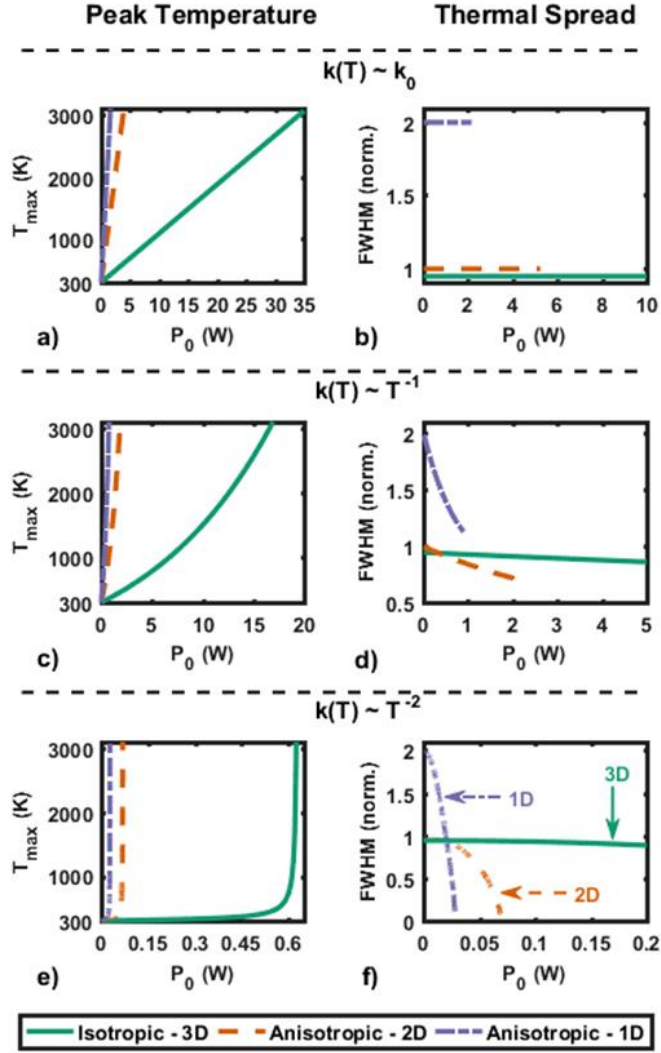

FIG. 5. Trends with input power. Peak temperature (a, c, e) and thermal spread normalized to the incident beam radius (b, d, f) for (a, b) constant, (c, d) $T^{-1}$, and (e, f) $T^{-2}$ thermal conductivity behaviors, as a function of the input power. The thermal spread is shown for the Y direction, which is along the nanotubes.

We define a heating efficiency as the portion of incident power that is converted to thermionic emission and incandescence (as opposed to dissipated through conduction). While, assuming a work function of 2.5 eV, the heating efficiency for 3D-$T^{-2}$ conductivity is only 1.95% at 3,000 K, it reaches 34.98% for 1D-$T^{-2}$ conductivity. The trends in peak temperature and temperature spread as a function of illuminated spot size, relative strength of thermionic emission and incandescence, heating efficiency, and results based on a temperature-dependent thermal conductivity measured for actual CNTs, are presented in Supplemental Material [25]. In particular, the model predicts that for a given optical intensity, a higher temperature is achieved with an increase in the size of the illuminating spot (Fig. S2). This comes from the fact that intensity depends on the inverse square of the spot diameter, whereas the primary heat conduction cross-sectional area (which is perpendicular to the nanotubes axis) is linearly proportional to the spot diameter, meaning that heat conduction along the nanotubes becomes relatively more difficult for wider spots. At the nano/meso scales, this may be viewed as follows: The radiation from each nanotube is partially reabsorbed by the adjacent nanotubes, such that each would absorb a higher effective power than if it were isolated. Therefore, starting from one nanotube and gradually placing additional ones adjacent to it, an increase in the temperature of the illuminated spot is expected. This may partially explain why experiments on optically heated individual CNTs [16, 17] have yielded much lower temperature levels compared to those on CNT forests [9].

Finally, we note that anisotropic thermal conductivity is not limited to 1D systems. For instance, black phosphorous also exhibits anisotropic thermal conductivity even within its basal plane [44]; its thermal conductivity also shows a peak at around 100 K and decays well past room temperature [45]. Therefore, the approach and model described here are not limited to CNTs. We also note that we have assumed the incandescence to follow Planck's law. While CNT forests are near-ideal black materials as mentioned before, Heat Trap may represent a non-equilibrium condition. Among other effects, the electronic and lattice temperatures may be different, as has also been observed in graphite and van der Waals heterostructures [46-48], and deviations from the black body radiation law are possible. However, prior works have shown good agreement with the black body radiation law for CNTs [49, 50]. Our ongoing work nonetheless involves incandescence spectroscopy over a broad range to elucidate possible deviations from Planck's law. We also point out that extremely large temperature gradients have also been observed in other low-dimensional systems such as silicon and aluminum wires with sub-micron diameters [51, 52], which further suggests that such heat localization may be a general property arising from low-dimensionality, rather than being limited to carbon nanotubes. We should also emphasize that an overall low thermal conductivity would also obviously lead to some degree of heat localization. (Given the porous and defective nature of the CNT forest and the inter-nanotube entanglements thereof, it is expected to have a lower thermal conductivity than pristine nanotubes.) Other published mechanisms such as Anderson localization or discrete breathers [53, 54] could also play a role in limiting heat flow. We thus do not claim that our proposed mechanism is a necessary condition for the Heat Trap effect, but that it is a sufficient one for it. It is also worth noting that, while we considered optical illumination for heating, similar localization effects might take place even if the input power is delivered through other means such as Joule heating [55-57].

**Summary**

In summary, we have demonstrated that material dimensionality/anisotropy, combined with the fact that thermal conductivity decreases with temperature, can result in strong localization of heat with extremely high temperature gradients in a direction where thermal conductivity is intrinsically high. This represents a mechanism for controlling heat flow in conductors, with significant implications: it means that one can maintain a high temperature difference between two points that are electrically connected, or create localized electron sources using low excitation powers. These could lead to breakthroughs in thermoelectricity and thermionics for energy conversion and a variety of other applications.

### Acknowledgments

We acknowledge financial support from the Natural Sciences and Engineering Research Council of Canada (SPG-P 478867, RGPIN-2017-04608, RGPAS-2017-507958), the Canada Foundation for Innovation, the British Columbia Knowledge Development Fund, and the Peter Wall Institute for Advanced Studies. This research was undertaken thanks in part to funding from the Canada First Research Excellence Fund, Quantum Materials and Future Technologies Program.

# Heat localization through reduced dimensionality
## *Supplemental Material*

Mike Chang,[1,4] Harrison D. E. Fan,[1,4] Mokter M. Chowdhury, [1,4] George A. Sawatzky,[2,3,4] Alireza Nojeh[1,4*]

[1]Department of Electrical and Computer Engineering, [2]Department of Physics and Astronomy
[3]Department of Chemistry, [4]Quantum Matter Institute
University of British Columbia, Vancouver, British Columbia, Canada
* Corresponding author: alireza.nojeh@ubc.ca

## I.     Methods

  Assume that thermal energy is delivered to a spot on the surface of a material (Fig. S1 (a)). The equation for conservation of energy is

$$-\nabla \cdot \vec{q} + Q_0 = \rho C_p \frac{\partial T}{\partial t},$$ (1)

where $\rho$ and $C_p$ are the material's density and specific heat, respectively, and $\vec{q}$ is the heat diffusion flux given by Fourier's law,

$$\vec{q}_i = -\sum_j k_{ij} \frac{\partial T}{\partial j} , \ i,j = x, y, z ,$$ (2)

where $k_{ij}$ are the elements of the thermal conductivity tensor of an anisotropic medium; $Q_0$ is the volume density of the thermal power delivered to the material. Steady-state solutions to the above equation have been found for systems under various conditions in the past [1–5].

  We assume that the material is generally anisotropic with the axes of anisotropy lying in the x, y and z directions. We further assume that the temperature dependence of thermal conductivity is the same for all three directions. The thermal conductivity tensor is then

$$k = \begin{bmatrix} \alpha & 0 & 0 \\ 0 & \beta & 0 \\ 0 & 0 & \gamma \end{bmatrix} k(T) ,$$ (3)

where $\alpha, \beta, \gamma$ are the thermal anisotropicity factors associated with the three directions and $k(T)$ represents the common temperature dependence.

The formulae chosen for $k(T)$ include three cases as shown in Fig. S1(b): one constant with temperature, the second one having a $T$ term as the dominant term at high temperatures in the denominator, and the third one having a $T^2$ term as the dominant term at high temperatures in the denominator. All three cases share the value of 100 W/mK at room temperature to represent a good conductor and allow a fair comparison among the three cases. We put no particular emphasis on the exact form of $k(T)$ and our goal is to simply show, qualitatively, that a higher-order drop with temperature leads to greater confinement. Nonetheless, the third case has been loosely inspired by the behavior reported for carbon nanotubes in the literature (see equation (10) further below), and the form of the second case has been chosen for consistency with the third case. The constants in the formulae have been chosen such that the thermal conductivity of the third case drops to approximately 0.01 W/mK (corresponding to a very good insulator) along the high-conductivity axis at 3,000 K (the upper bound of the temperature range we have considered in the manuscript); in anisotropic cases, the conductivity value along the other axes becomes essentially negligible in the high-temperature limit, helping to emphasize the anisotropicity. The plots of the three different cases in Fig. S1(b) show their distinct behaviors.

Without loss of conceptual generality, we assume that heating is accomplished through illumination by a Gaussian beam of light with power $P_0$.

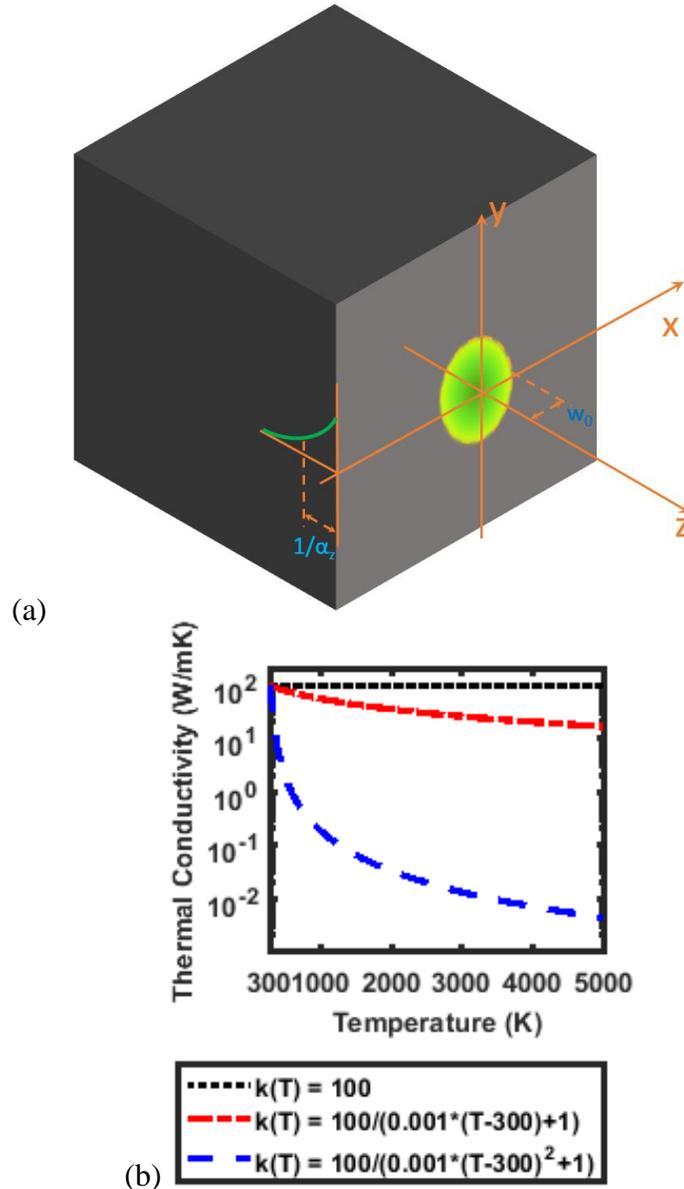

(a)

(b)

FIG. S1 (a) Schematic of the model, consisting of optical energy delivered to the surface with an exponentially decaying tail into the bulk due to absorption by the material. The boundaries are held at room temperature at infinity (Dirichlet boundary condition) except for the front surface (Neumann boundary condition). (b) The three behaviours of the temperature dependence of thermal conductivity used in this study.

The non-linearity introduced by the temperature dependence of thermal conductivity can be dealt with through the Kirchhoff transform [1,6–9] by introducing a so-called linear temperature as

$$\phi(x,y,z,t) = \frac{1}{k_0} \int_{T_0}^{T(x,y,z,t)} k(\tau)\,\partial\tau, \tag{4}$$

$$k_0 = k(T_0),$$

where $T$ is true temperature and $T_0$ its value infinitely far from the point of illumination. With a coordinate transformation, the heat equation is simplified to

$$k_0 \left[ -\nabla^2{}_{x',y',z'} + \frac{\rho C_p(T)}{k(T)} \frac{\partial}{\partial t} \right] \phi(x',y',z',t)$$

$$= \frac{2P_0 \alpha_z}{\pi w_0{}^2} e^{-2\frac{\alpha x'^2}{w_0{}^2} - 2\frac{\beta y'^2}{w_0{}^2}} e^{-\alpha_z \sqrt{\gamma} z'}, \tag{5}$$

where $x'$, $y'$, $z'$ are the coordinates scaled by the square roots of the corresponding thermal anisotropicity factors, and $\alpha_z$ and $w_0$ are the optical absorption coefficient of the medium (which we take to be independent of temperature) and the $e^{-2}$ width of the Gaussian beam, respectively. The steady-state solution to an equation similar to Eq. (5), but with no z dependence (assuming a Dirac delta absorption at the surface) has been derived by Lu [1] through the Green's function method. Following the same approach, the solution for the case of exponentially decaying optical absorption can be derived as

$$\phi(x,y,z) = \frac{\sqrt{2}P_0}{w_0} \frac{1}{\pi^{\frac{3}{2}}(\alpha\beta\gamma^2)^{\frac{1}{4}}k_0} \int_0^\infty f(u,x,y)g(u,z)du, \tag{6}$$

where the spatial distribution functions are given by

$$f(u,x,y) = \frac{e^{-\frac{2}{w_0{}^2}\left(\frac{x^2}{\sqrt{\frac{\alpha}{\beta}}u^2+1} + \frac{y^2}{\sqrt{\frac{\beta}{\alpha}}u^2+1}\right)}}{\sqrt{\left(u^2+\sqrt{\frac{\beta}{\alpha}}\right)\left(u^2+\sqrt{\frac{\alpha}{\beta}}\right)}}, \tag{7}$$

and

$$g(u,z) = \sqrt{\pi}(\delta_0 u)erfc\left(\delta_0 u - \frac{\alpha_z z}{2}\frac{1}{\delta_0 u}\right)e^{(\delta_0 u)^2 - \alpha_z z}, \tag{8}$$

$$\delta_0 = \frac{\alpha_z w_0}{2\sqrt{2}}\left(\frac{\gamma^2}{\alpha\beta}\right)^{1/4}. \tag{9}$$

Note that if the Gaussian width of the incident beam is much larger than the optical penetration depth, or the thermal conductivity along the depth is much higher than those along the surface directions, the parameter $\delta_0$ will have a large value and the solution will approach that for the case of complete optical absorption at the surface.

Two immediate and critical observations can be made from Eqs. (6) and (4): First, the temperature distribution along any direction is inversely proportional to $(\alpha\beta\gamma^2)^{1/4}$, which involves the thermal conductivities in all three directions. This means that having low thermal conductivities in two directions can help increase the temperature along a third direction, even if thermal conductivity along the latter is not low in itself. Second, the profile of the induced linear temperature simply scales with total input power; however, if thermal conductivity decreases with temperature (as is common for crystals at high temperatures [10]), an increase in linear temperature means a stronger corresponding increase in true temperature. In other words, an increase in input power will lead to a non-linearly amplified increase in true temperature. This effect will be stronger at higher temperatures, causing the temperature profile to be pinched spatially around the center of the illuminated spot. The above two effects form the foundation of our proposed mechanism for maintaining a high temperature gradient in a conductor.

In order to add radiative loss to the model, the incandescence (black-body radiation) intensity is calculated as a function of x and y using the Stefan-Boltzmann law and integrated over the surface. This is carried out in a self-consistent loop where the temperature distribution is first calculated using the model described above for a given input power, the incandescence loss is then calculated and the effective input power reduced accordingly, the temperature distribution is then recalculated, and so on, until convergence. Where applicable, energy loss through thermionic electron emission is also incorporated into the self-consistent loop in a similar manner, based on the Richardson equation [11] (with $A_G$=120 A/cm$^2$K$^2$).

We assume that a circular Gaussian beam with an e$^{-2}$ diameter of 100 μm is incident on the surface; we take the optical absorption coefficient to be 2.9 μm$^{-1}$ [12], corresponding to a penetration depth of 344.8 nm. To calculate incandescence loss, we assume an emissivity of 1 for all cases. This assumption of ideal black-body emission is motivated by the behaviour of carbon nanotube forests which are nearly-perfect dark materials over a broad spectral range [13,14], and is here used in all cases of anisotropy for simplicity and to enable a fair comparison. A lower value of emissivity would lead to different values of temperatures, but would not affect our conclusions qualitatively.

## II.     Effect of illuminated spot diameter

Figure S2 illustrates the peak temperature and FWHM (in the y direction) of the temperature profile as a function of the diameter of the illuminating beam spot, while keeping the input optical intensity constant throughout each curve by adjusting the total input power as the beam radius increases. It is interesting to see that, for a fixed optical intensity, a higher temperature is achieved for a higher spot size. This may be explained as follows: as the spot radius is increased, the amount of input power required to keep the intensity constant, increases quadratically; however, the amount of heat conduction away from the illuminated spot does not rise at the same rate. This effect is most pronounced in the 1D case, where heat conduction happens along the y axis, and thus the cross-sectional area of heat conduction is determined by the penetration depth of the temperature profile multiplied by the width of the profile in the x direction. Doubling the spot diameter only doubles the heat dissipation cross section, but corresponds to a four-fold increase in input power for the same optical intensity. This lower level of conductive loss relative to input power, leads to higher temperature. For constant thermal conductivity and T$^{-1}$, the temperature distribution also broadens correspondingly. Interestingly, for T$^{-2}$ and 1D, the temperature profile also initially broadens as the illuminated spot radius is increased, but only up until a tipping point beyond which any further increase in the optical spread translates to a diminishing thermal spread. This is a result of the aforementioned effect of a greatly declining thermal conductivity on thermal spread at elevated temperatures: the rate of increase of the peak temperature with the optical spot radius becomes higher than the rate of broadening of the temperature distribution, leading to a decreasing FWHM at larger optical spot sizes.

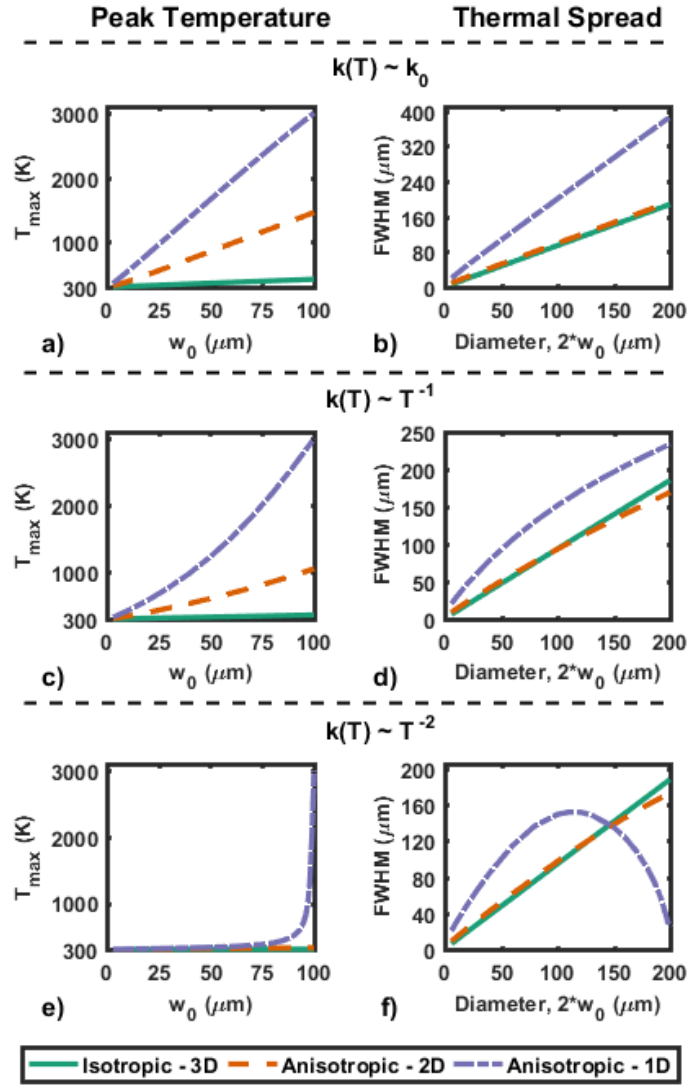

FIG. S2. Peak temperature (a, c, e) and thermal spread (b, d, f) for (a, b) constant, (c, d) $T^{-1}$, and (e, f) $T^{-2}$ thermal conductivity behaviors, as a function of the input optical beam radius. The thermal spread is shown for the y direction (direction of high thermal conductivity). The optical intensity is kept constant throughout each curve by adjusting the input power as the beam radius is increased.

### III.   Relative strength of different heat loss mechanisms, and heating efficiency

Figures S3 and S4 show the contributions of the various power loss mechanisms for the case of $T^{-2}$ thermal conductivity. (Figure S4 is the same as Fig. S3, but with a more limited range of the axes, in order to show the various contributions better in a practical temperature range.) Two material workfunctions of 4.6 eV and 2.5 eV, typical values respectively for carbon nanotubes and low-workfunction thermionic cathodes, were chosen. At low temperatures, conduction dominates. At high temperatures, radiation and thermionic emission become increasingly important, and lead to a slowing down of the rate of temperature increase with input power. The effect is strongest in the 1D case, where the highest proportion of incident power contributing to thermionic emission is attained. This suggests that a 1D material provides the best option for a thermionic cathode for energy conversion and vacuum electronic applications.

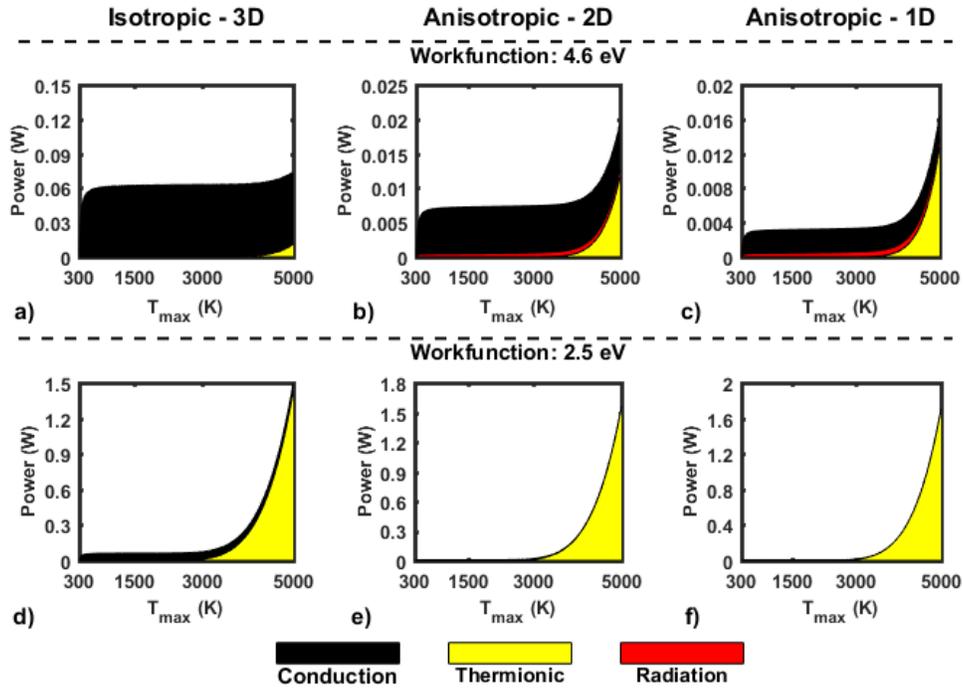

FIG. S3. Distribution of power loss among conduction, radiation and thermionic emission mechanisms as a function of peak temperature. Thermal conductivity has been assumed to behave as $T^{-2}$ but with a peak value of 10 W/mK (different from that used in the main text) in order to strengthen the relative strength of the thermionic component for demonstration purposes (this point is further illustrated in Fig. S4). The two rows correspond to different values of workfunction, with the lower-workfunction case clearly showing a strong thermionic emission component. Each column corresponds to a different case of anisotropy. Conductive power loss was calculated by deducting the radiation and thermionic emission loss powers from the total input power.

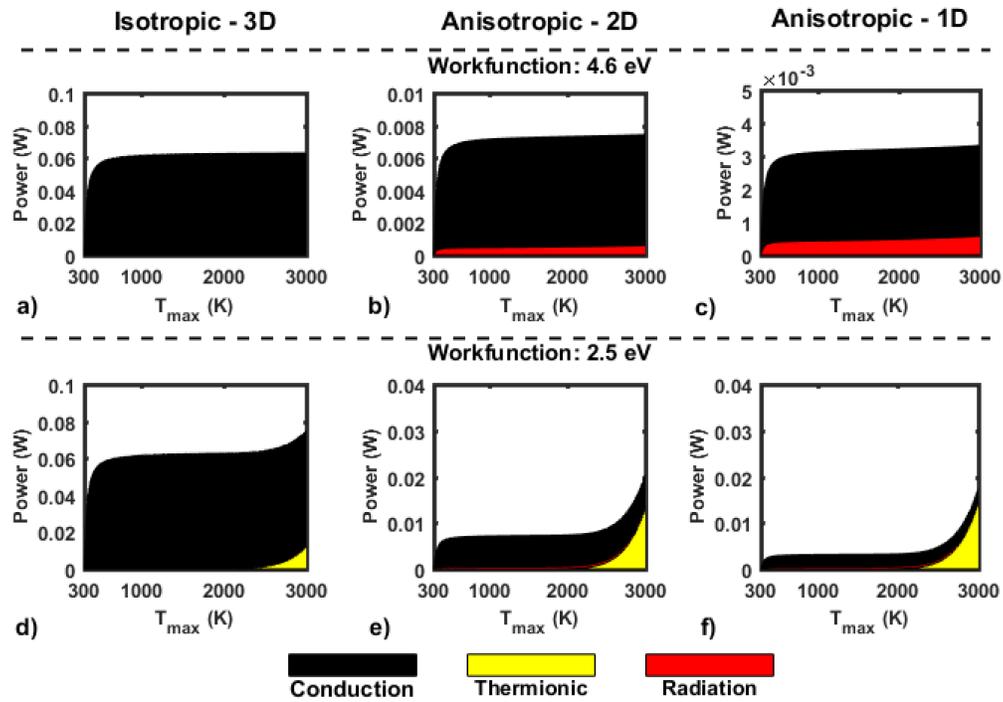

FIG. S4. Same as Fig. S3, but with smaller ranges of the axes to show the trends better at lower temperatures.

We define a heating efficiency as the ratio of the amount of power that goes into incandescence plus thermionic emission, over total power. Figure S5(a) shows the heating efficiency versus input power (assuming $T^{-2}$ thermal conductivity), clearly showing the superiority of the 1D case, where efficiencies of 1.59 % (at 27.26 mW of input power, corresponding to a temperature of 1,000 K) and 34.98 % (at 42.17 mW of input power, corresponding to a temperature of 3,000 K) are achieved for a material with a room temperature thermal conductivity of 100 W/mK. The situation would expectedly improve even further if a material with a lower room temperature conductivity of 10 W/mK is used, leading to a heating efficiency of 13.95 % (at 3.12 mW, 1,000 K) and 84.32 % (at 17.49 mW, 3,000 K) (Fig. S5(b)).

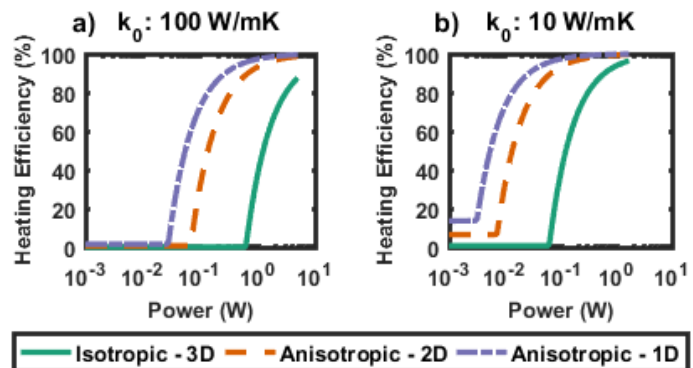

FIG. S5. Effect of dimensionality on heating efficiency as a function of input power for a material with a workfunction of 2.5 eV and a room temperature thermal conductivity of 100 W/mK (a) or 10 W/mK (b). Thermal conductivity has been assumed to behave as $T^{-2}$ in all cases.

## IV.    Modeling a CNT forest with an empirically obtained behaviour of thermal conductivity

A potential physical realization of the 1D case is a CNT forest. Such structures can be made from single- and multi-walled nanotubes, which have diameters on the order of a nanometer to a few tens of nanometers. The inter-nanotube spacing in the forest can be several tens of nanometers. The structure thus consists mostly of empty space (up to 97-98%). The nanotubes have long-range alignment. There is also a significant level of inter-nanotube contact due to local deviations from a straight line. Nonetheless, the forest represents an example of a highly anisotropic material, with a thermal anisotropicity of 72 reported in ref. [15]; for a hypothetical array of perfectly straight nanotubes, an even higher anisotropicity factor is expected. (For multi-layer graphene, the anisotropicity ratio is roughly two orders of magnitude [10].) Such forests can be grown to macroscopic dimensions, and also have near-perfect optical absorption over a broad range of wavelengths [13,14]. There have been extensive studies on the synthesis of this material in regards to alignment, density, growth rate and macroscopic dimensions using catalytic chemical vapour deposition. Depending on the level of defects and inter-nanotube entanglement in the forest, its thermal conductivity even along the nanotubes may be significantly lower than that of a pristine, isolated nanotube. However, we will show that even if each nanotube retains its high individual thermal conductivity, the mechanism described in this work will lead to significant localized heating. To this end, we will assume that the thermal conductivity of each nanotube obeys the empirical temperature dependence reported for an individual single-walled carbon nanotube [16]:

$$k(T; L[\mu m]) = \frac{1}{(3.7 \times 10^{-7})T + (9.7 \times 10^{-10})T^2 + \dfrac{9.3}{T^2}\left(1 + \dfrac{0.5}{L[\mu m]}\right)} \qquad (10)$$

, where $L[\mu m]$ is the length of the nanotube in microns to account for the boundary effect when the tube length is on the order of the mean free path of phonons in carbon nanotubes.

Figure S6 shows the resulting temperature distribution, revealing a peak temperature of 3,010 K and temperature gradients in excess of 61.6 K/µm and 108.6 K/µm along the y and x axes, respectively, for an input optical power of 245 mW and corresponding linear intensity of 4.9 kW/m and aerial intensity of 31.2 W/mm² (this intensity is much lower than what is required to heat regular bulk conductors), confirming that the mechanism revealed in this paper is indeed applicable to real physical systems, as further supported by our experimental observation of the Heat Trap effect [17].

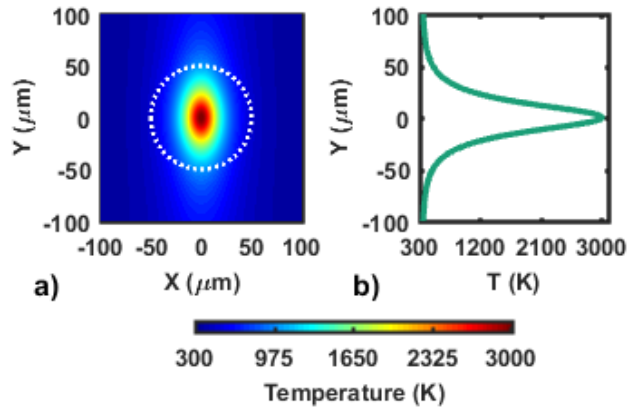

FIG. S6. Laser-induced temperature profile for a hypothetical array of 1 millimeter-tall, perfectly-aligned single-walled carbon nanotubes with a fill-factor of 2% and anisotropicity of 72 between the longitudinal and transverse directions. (a) is the temperature distribution and (b) its cross section along the y axis (nanotube length). The dashed circle represents the illuminating beam spot.